\documentclass[twocolumn]{aastex63}

\received{January 16, 2020}
\revised{April 16, 2020}
\accepted{April 17, 2020}
\published{May 18, 2020}

\shorttitle{Speckle interferometry of nearby multiple stars}
\shortauthors{Mitrofanova et al.}

\begin{document}

\title{Speckle Interferometry of Nearby Multiple Stars: 2007-2019 Positional Measurements and Orbits of Eight Objects.}

\correspondingauthor{Arina Mitrofanova}
\email{arishka670a@mail.ru}

\author[0000-0002-2305-5499]{A. Mitrofanova}
\affiliation{Special Astrophysical Observatory,\\
369167 Nizhnij Arkhyz, Russia}

\author{V. Dyachenko}
\affiliation{Special Astrophysical Observatory,\\
	369167 Nizhnij Arkhyz, Russia}

\author{A. Beskakotov}
\affiliation{Special Astrophysical Observatory,\\
	369167 Nizhnij Arkhyz, Russia}

\author{Yu. Balega}
\affiliation{Special Astrophysical Observatory,\\
	369167 Nizhnij Arkhyz, Russia}

\author{A. Maksimov}
\affiliation{Special Astrophysical Observatory,\\
	369167 Nizhnij Arkhyz, Russia}

\author{D. Rastegaev}
\affiliation{Special Astrophysical Observatory,\\
	369167 Nizhnij Arkhyz, Russia}

\author{S. Komarinsky}
\affiliation{Special Astrophysical Observatory,\\
	369167 Nizhnij Arkhyz, Russia}

\begin{abstract}

The orbits of 8 systems with low-mass components (HIP~14524, HIP~16025, HIP~28671, HIP~46199, HIP~47791, HIP~60444, HIP~61100 and HIP~73085) are presented. Speckle interferometric data were obtained at the 6 m Big Alt-azimuth Special Astrophysical Observatory of the Russian Academy of Sciences (BTA SAO RAS) from 2007 to 2019. New data, together with measures already in the literature, made it possible to improve upon previous orbital solutions in six cases and to construct orbits for the first time in the two remaining cases (HIP~14524 and HIP~60444). Mass sums are obtained using both Hipparcos and Gaia parallaxes, and a comparison with previously published values is made. Using the Worley \& Heintz criteria, the classification of the orbits constructed is carried out.

\end{abstract}

\keywords{observational astronomy --- astronomical techniques --- interferometry --- speckle interferometry, stellar astronomy --- stellar types --- low mass stars, interferometric binary}

\section{Introduction} \label{sec:intro}

Most resolvable binaries consist of late-type main-sequence stars. The study of such objects using the speckle interferometry \citep{lab70} allows for the determination of their fundamental parameters (the mass sum of the components and, if there are additional data, their individual masses and spectral types), the compilation of statistical samples according to the obtained orbital parameters, and an increase in the known multiplicity of systems previously defined as single/double. The objects in this study are stars resolved by the Hipparcos satellite \citep{hip}: HIP~14524, HIP~16025, HIP~28671, HIP~46199, HIP~60444 and HIP~73085. Two objects were studied in previous papers, HIP~61100 by \citet{bali13} and HIP~47791 by \citet{mcal93}. 

One of the main aims of speckle interferometry as an observational method is to obtain images of binary stars at angular scales from the atmospheric seeing size (about 1\arcsec-2\arcsec) to the diffraction limit (as small as 0\farcs02 for the 6 m telescope at a wavelength of 500 nm). Among the pairs available at such angular separations, there are systems with periods from $\sim$1 to $\sim$100 yr. Long-term monitoring of binary and multiple systems allows for the calculation of high-precision orbital elements, and therefore direct determination of the masses of stars and the study of stellar evolution. 

Currently, studies of binary systems by speckle interferometry are carried out by several teams worldwide (e.g. \citet{mas01, mas18, bali13, hart15, orl15, doc17, doc19, hor17, hor19, men18, tok18, tok19, tok19a, tok19b}) on medium and large telescopes (from 1.5 to 8.1 m). However, despite the rather large amount of studies already carried out, many systems do not have orbital solutions at all or they are not accurate. The results of the Gaia mission \citep{gplx} will significantly improve our knowledge about the orbits of such objects, and therefore about the fundamental parameters of stars. It should be noted that, for many binaries, the observations of this satellite will cover only a part of the orbital arc (which will be very small for objects with long periods, even taking into account observations in phases close to the periastron). At the same time, when observations of the Gaia mission correspond to phases far from the passage of the periastron, the constructed orbital solutions will probably be ambiguous. In this case, long-term ground-based monitoring of binary systems will allow for the construction of the correct orbital solution.

The monitoring of the eight objects that are the subject of this paper has been carried out by the group of high-resolution methods in astronomy at the Special Astrophysical Observatory of the Russian Academy of Sciences (SAO RAS) since the late 1990s. Some of the astrometric results have already been published earlier by \citet{bali02, bali04, bali06, bali07}, \citet{rast07, rast08},  and \citet{bali13} and are presented in Table \ref{tab1}. This work is dedicated to the further study of these systems and the construction of their orbits. Despite the fact that orbital solutions have already been published for some objects (Table \ref{tab2}), they should be improved due to a lack of observational data and long orbital periods. Section 2 provides a description of speckle interferometric observations and data reduction results. Section 3 is dedicated to the construction of orbits and the analysis of the orbital solutions obtained.

\section{Observations and Data Reduction} \label{sec:observations}

\begin{figure}
	\plotone{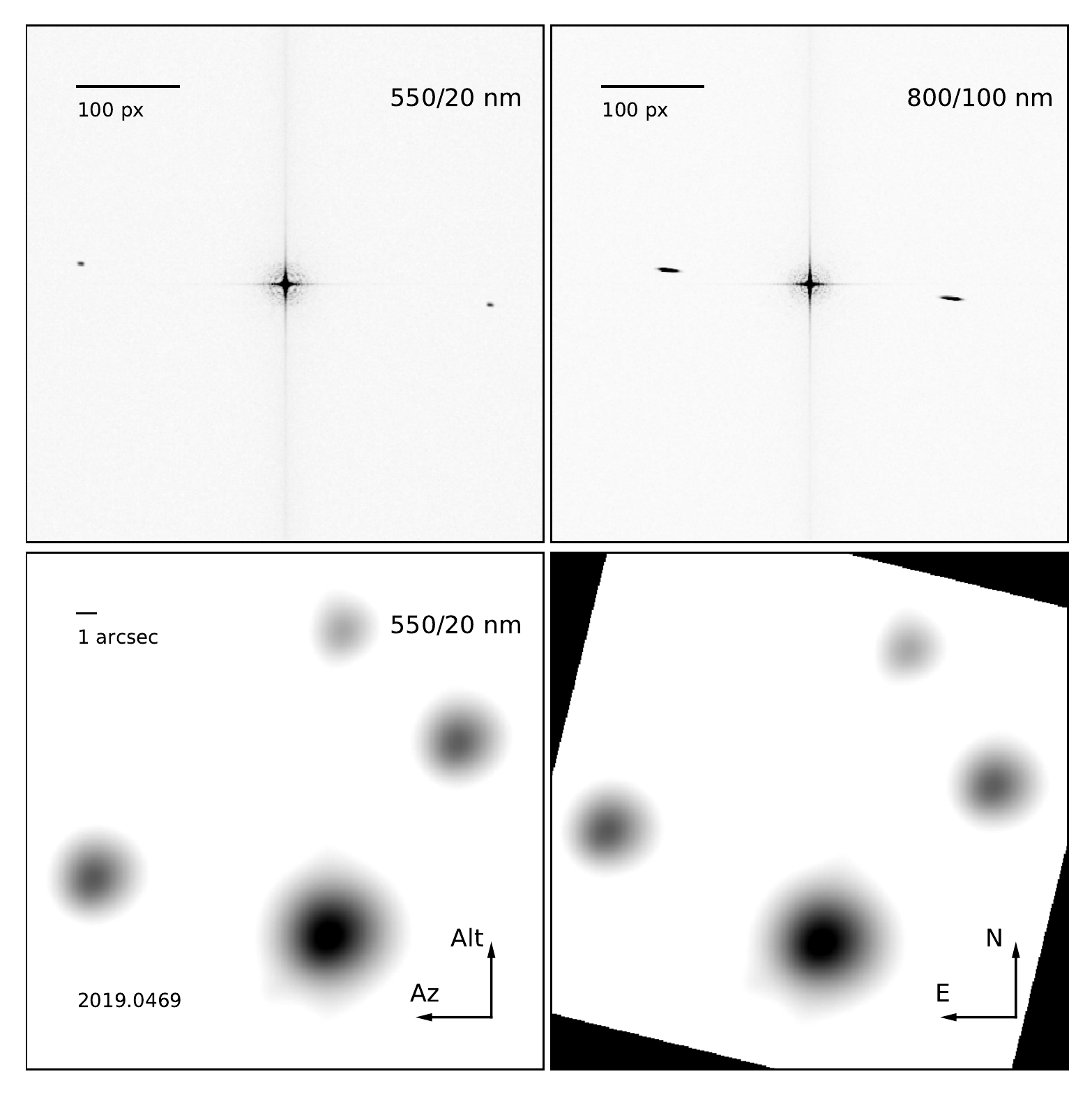}
	\caption{Top: examples of power spectra of a series of interference fringes from the two-hole mask when observed in 550/20 nm (left) and 800/100 nm (right) bands. Bottom: the image of the $\theta^{1}$ Ori (Trapezium Cluster) averaged over 500 frames in the horizontal coordinate system (left) and after reduction to the equatorial coordinate system (right).  \label{fig1}}
\end{figure}

Speckle interferometric observations were carried out in 2007-2019 at the Big Telescope Alt-azimuth (BTA) of the SAO RAS using a speckle interferometer \citep{maks09} based on EMCCD detectors, namely the PhotonMAX-512B (until 2010), Andor iXon+ X-3974 (2010-2014), and Andor iXon Ultra 897 (since 2015). The observation log is shown in Table \ref{tab1}, most of the speckle images were obtained under good weather conditions with seeing 1\arcsec-2\arcsec. Speckle interferograms were recorded with an exposure time of 20 milliseconds and a standard series of 1940 (until 2010) and 2000 images for each object. The following interference filters were used (central wavelength $\lambda$/bandpass $\Delta \lambda$): 550/20 nm, 600/40 nm, 800/100 nm and 900/80 nm. High measurement accuracy is due to the simultaneous use of the following methods for calibration: 
\begin{itemize}
	\item For the position angle and the angular separation calibration we use a two-hole mask installed in the converging $F/4$ beam at the distance of 3995 mm from the focal plane. Physically, this mask represents the Hartmann diaphragm at the entrance to the BTA primary focus cage, in which only two of the 260 holes are left open. The diaphragm is usually used to take pre-focal and extra-focal images when studying the surface shape of the telescope's main mirror. The distance between the 15 mm diameter holes is $833 \pm 1$ mm. The holes are symmetric with respect to the telescope axis. Light from a bright star passes through the holes and produces a set of fringes in the focal plane of the mirror. The orientation of the fringes, defined by the angle of the line connecting the holes, is always fixed and can be used for the calibration of the installation angle of the camera. The linear spacing between the fringes is determined by the distance between the calibration holes and the focal length of the telescope and is given by
	
	\begin{equation}
	M = \frac{360*60*60}{2*\pi*F}*\frac{\lambda*L}{d}/\frac{N}{\rho(\lambda)},
	\label{eq1}
	\end{equation}
	
	where M is scale in arcseconds per pixel, F is telescope focal length in meters, $\lambda$ is the wavelength (m), L is the distance from the entrance of the telescope prime cage to the focal plane (m), d is the distance between the holes on the cover panels of the prime focus cage (m), N is the number of pixels in the original image, and $\rho(\lambda)$ is the measured distance in pixels between the peaks in the power spectra series of the mask.
	
	The main source of uncertainty in determining the scale is the value of the focal length of the mirror. Workshop measurements of the radius of the curvature of the mirror, made in different years and in different zones along the radius of the mirror, as well as measurements of the focus position using images of the Hartmann diaphragm on a telescope, differ by tens of millimeters. The average value of the focus of the BTA mirror adopted by us is $24025 \pm 15$ mm. In the projection onto the mirror surface, the calibration holes are spaced 4994 mm apart, and their diameter corresponds to 90 mm.
	
	\item Calibration of the scale and angle is performed on a bright star observed with good quality atmospheric images. Under poor atmospheric conditions (seeing ~ 2\arcsec - 3\arcsec) the phase differences of the wave fronts from the two apertures are too large, the images of the star do not overlap most of the time, and the contrast of the accumulated interference pattern is low. In the frequency plane, interference fringes correspond to a pair of elongated spots (Figure \ref{fig1}, top panel). Their length is determined by the spectral band of the used interference filter: the narrower the filter, the more accurate the measurement of scale. We are currently scaling with relative accuracy better than 0.05 mas per pixel (this value corresponds to the least precise determinations of the scale among the 2007-2019 measurements in different bandpasses). The orientation of the speckle interferometer is also related to the mask. The value of the position angle in the equatorial coordinate system is calculated from the parallactic angle of the star at the time of observation. We have been calibrating the angle in the sky for many years from the images of the $\theta^{1}$ Ori (Trapezium Cluster) using a x2.5 microlens (corresponding field of view is 28\farcs2), shown in the bottom panel of Figure \ref{fig1}. The position angles between the components of which were determined by hundreds of observers using a variety of methods, so they can be considered very reliable (e.g. \citet{oli13}).
\end{itemize}

\begin{figure*}
	\plotone{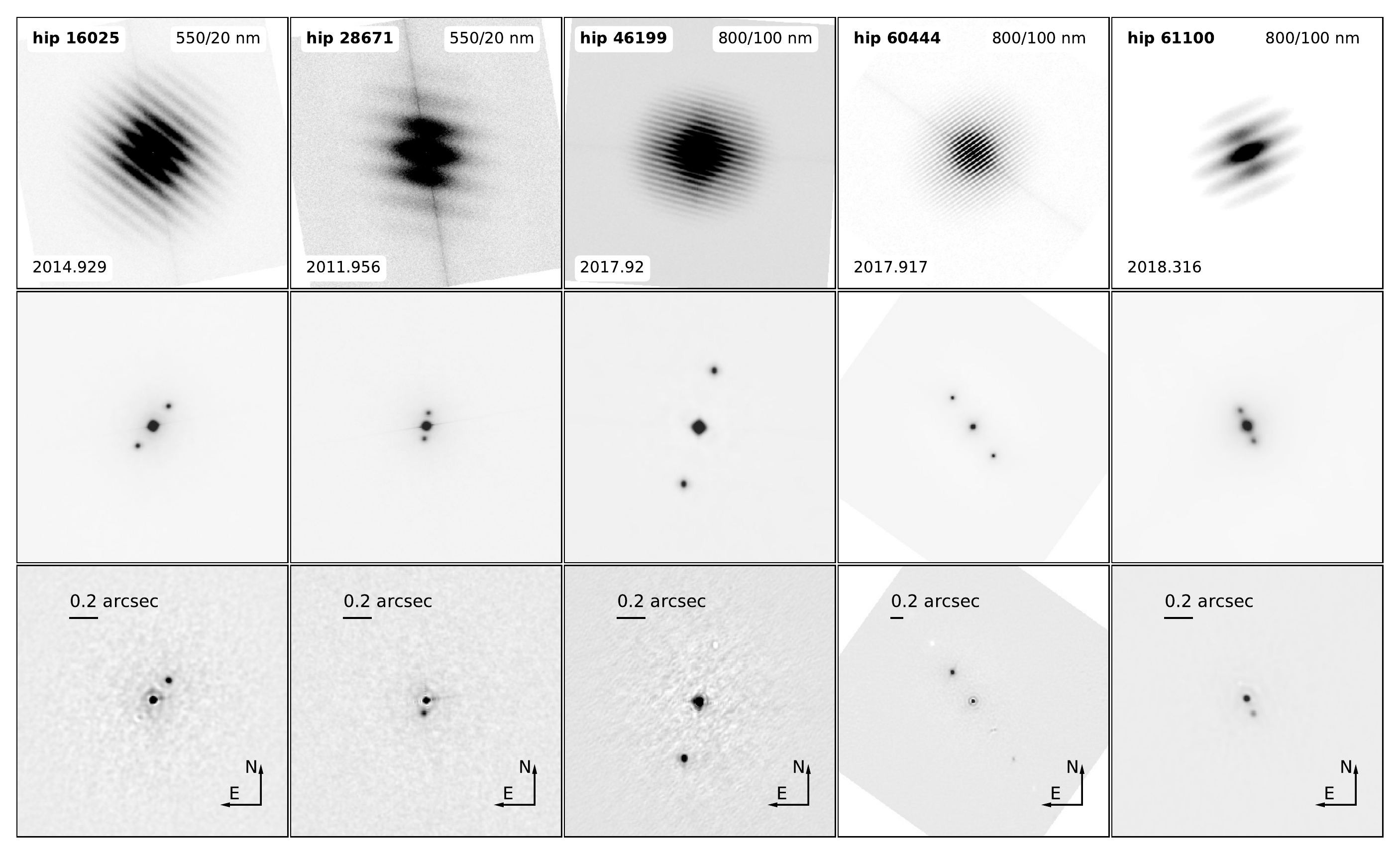}
	\caption{Examples of power spectra, corresponding autocorrelation functions, and reconstructed images for the objects studied.  \label{fig2}}
\end{figure*}

Positional parameters and magnitude differences were determined through the analysis of the power spectrum and the autocorrelation function of the speckle interferometric series described in \citet{bali02} and \citet{pluz05}. The reconstruction of the position of the secondary was carried out by the method of bispectral analysis \citep{lohm83}. Figure \ref{fig2} shows examples of power spectra, corresponding autocorrelation functions, and reconstructed images for the objects studied. 

Table \ref{tab1} presents the positional parameters of the systems under study and the magnitude differences, known from the literature and obtained in this study. The columns are Hipparcos number, epoch of observation in fractions of Besselian year, telescope, bandpass or $\lambda$/$\Delta\lambda$, $\theta$ and $\sigma_{\theta}$ are the position angle of the secondary relative to the primary and its error, $\rho$ and $\sigma_{\rho}$ are the separation between the two stars and its error, $\Delta m$ and $\sigma_{\Delta m}$ are the magnitude difference and its error, and references. The analysis of the magnitude differences from the data obtained in different epochs shows that the average population standard deviation of new measurements is about 0.1 mag.

\begin{deluxetable*}{lllllllllll}
	\tablenum{1}
	\tablecaption{The positional parameters of the objects studied and the magnitude differences between components. \label{tab1}}
	\tablewidth{0pt}
	\tablehead{
		\colhead{HIP} & \colhead{Epoch, yr} & \colhead{Telescope} & \colhead{$\lambda$ / $\Delta\lambda$, nm}  & \colhead{$\theta\degr$} & \colhead{$\sigma_{\theta}\degr$} & \colhead{$\rho$, mas} & \colhead{$\sigma_{\rho}$, mas} & \colhead{$\Delta m$, mag} & \colhead{$\sigma_{\Delta m}$, mag} & \colhead{Reference}
	}
	\decimalcolnumbers
	\startdata
	14524  & 2008.712 & BTA & 550/20 & 303.1 & 0.1 & 327 & 1 & 0.82 &  & This work \\
	& 2009.9107 & BTA & 800/100 & 298.6 & 0.1 & 316 & 1 & 1.13 &  & This work \\
	& 2011.9505 & BTA & 800/100 & 288.5 & 0.2 & 299 & 1 & 0.22 &  & This work \\
	& 2015.9151 & BTA & 800/100 & 261.4 & 0.1 & 254 & 1 & 0.20 &  & This work \\
	& 2016.8869 & BTA & 800/100 & 252.6 & 0.1 & 235 & 1 & 0.28 &  & This work \\
	& 2017.7715 & BTA & 800/100 & 243.7 & 0.1 & 215 & 1 & 0.43 &  & This work \\
	& 2019.0495 & BTA & 800/100 & 226.8 & 0.1 & 174 & 1 & 0.37 &  & This work \\
	\enddata
	\tablecomments{This table is available in its entirety in machine-readable form. The full version of Table \ref{tab1} lists the parameters published earlier by \citet{hip}, \citet{bali02, bali04, bali06, bali07, bali13}, \citet{hor02, hor04, hor08, hor10, hor12, hor15, hor17}, \citet{tok14, tok15, tok16}, \citet{tok18}, \citet{mcal93}, \citet{hart00, hart12}, \citet{orl11}, \citet{sch16}, and \citet{rast07, rast08}.}
\end{deluxetable*}

\section{Orbit construction} \label{sec:orbit}

Preliminary estimates of the orbital parameters were calculated using the Monet method \citep{mon77}. The final orbital solutions were calculated using the ORBIT \citep{tok92} software package. Depending on the values of the residuals and deviations from the orbital solution, the corresponding weights were selected for each measurement (ones with large errors are set to smaller weights). When constructing the orbits, $180\degr$ ambiguities were found in the positions of measurements corresponding to previously published ones, which is probably due to inaccurate reconstruction of the position of the secondary or its absence. As a result, the values of the position angles of the following measurements were changed by $\pm 180\degr$ (in Table \ref{tab1}, the values of the position angles correspond to the published data):
\begin{itemize}
	\item HIP~14524: \citet{bali04}, \citet{bali06}, \citet{bali13}, and \citet{hor17};
	\item HIP~28671: 1991.25 \citet{hip} and 2010.9659 (both measurements) \citet{hart12};
	\item HIP~46199: \citet{bali04};
	\item HIP~47791: 1993.2025 \citet{mcal93};
	\item HIP~60444: 1999.8217 \citet{bali04};
	\item HIP~61100: 2002.2543 (both measurements), 2003.9258, 2005.2353, and 2005.2461 \citet{bali13};
	\item HIP~73085: 2014.2189 and 2014.4593 for 880/50 bandpass \citet{hor15}.
\end{itemize} 

It should be noted that all measurements presented in Table \ref{tab1} were used to construct orbits. Since positional parameters were obtained using different instruments and methods, they have different small systematic errors. Therefore, orbital residuals on $\rho$ and $\theta$, discussed below for each object, can be higher than the estimates of uncertainties presented in Table \ref{tab1}.

Table \ref{tab2} presents both our orbital parameters for the eight systems and those found in the literature. The columns give: the Hipparcos number, the orbital period in years, the epoch of periastron passage, the eccentricity, the semi-major axis in mas, the longitude of the ascending node, the argument of periastron, the inclination, and the reference for the calculation.

\begin{deluxetable*}{lllllllll}
	\tablenum{2}
	\tablecaption{Orbital parameters of the objects. \label{tab2}}
	\tablewidth{0pt}
	\tablehead{
		\colhead{HIP}  & \colhead{$P_{orb}$, year} & \colhead{$T_{0}$, year} & \colhead{$e$} & \colhead{$a$, mas} & \colhead{$\Omega$, $\degr$} &  \colhead{$\omega$, $\degr$} & \colhead{$i$, $\degr$} & \colhead{Reference}
	}
	\decimalcolnumbers
	\startdata
14524 & 33.6 & 2020.8 & 0.678 & 324.9 & 219.8 & 79.8 & 129.3 & This work \\
  & $\pm 0.7$ & $\pm 0.1$ & $\pm 0.032$ & $\pm 15.7$ & $\pm 1.7$ & $\pm 0.5$ & $\pm 3.3$ & \\ 
\hline
16025 & 54.695 & 2021.606 & 0.091 & 273.2 & 54.4 & 313.4 & 41.2 & \citet{cve16} \\  
  & $\pm0.881$ & $\pm 0.576$ & $\pm 0.007$ & $\pm 3.3$ & $\pm 1.1$ & $\pm 3.3$ & $\pm 0.7$ & \\  \cline{2-9}
  & 48.3 & 2020.9 & 0.158 & 257.1 & 56.2 & 325.1 & 41.9 & This work \\ 
  & $\pm 1.5$ & $\pm 0.9$ & $\pm 0.007$ & $\pm 4.1$ & $\pm 2.8$ & $\pm 12.1$ & $\pm 0.8$ & \\
\hline
28671 & 100 & 2044.6 & 0.93 & 449 & 51.5 & 80.2 & 95.9 & \citet{tok16a} \\  \cline{2-9}
  & 100.7 & 2039.0 & 0.955 & 567 & 49.8 & 80.4 & 94.6 & This work \\ 
  & $\pm 3$ & $\pm 0.5$ & $\pm 0.002$ & $\pm 46$ & $\pm 3.3$ & $\pm 0.5$ & $\pm 0.4$ & \\
\hline
46199 & 93.916 & 1995.977 & 0.835 & 582 & 140.6 & 263.4 & 76.9 & \citet{cve16} \\ 
  & $\pm 4.056$ & $\pm 4.423$ & $\pm 0.667$ & $\pm 32.7$ & $\pm 3.2$ & $\pm 23.9$ & $\pm 6.6$ & \\  \cline{2-9}
  & 111.8 & 1996.2 & 0.891 & 693.7 & 138.6 & 264.8 & 77.8 & This work \\ 
  & $\pm 4.8$ & $\pm 0.2$ & $\pm 0.008$ & $\pm 20.5$ & $\pm 0.5$ & $\pm 0.5$ & $\pm 0.4$ & \\
\hline
47791 & $21.724$ & $2000.621$ & 0.351 & 115.2 & 25.5 & 262.3 & 74.4 & \citet{cve16} \\  \cline{2-9}
  & 20.73 & 1999.78 & 0.373 & 114.1 & 22.2 & 259.7 & 76.3 & This work \\ 
  & $\pm 0.01$ & $\pm 0.01$ & $\pm 0.008$ & $\pm 0.7$ & $\pm 0.2$ & $\pm 0.1$ & $\pm 0.2$ & \\
\hline
60444 & 26.16 & 2000.78 & 0.633 & 602.3 & 24.2 & 113.0 & 99.3 & This work \\
  & $\pm 0.01$ & $\pm 0.01$ & $\pm 0.002$ & $\pm 3.3$ & $\pm 0.2$ & $\pm 0.07$ & $\pm 0.1$ & \\ 
\hline
61100 & $1284.37^{d}$ & 1988.96178 & 0.507 & 46.9 & 357.2 & 248.8 & 60.1 & \citet{are00} \\
  & $\pm 2.25^{d}$ & $\pm 0.01779$ & $\pm 0.015$ & $\pm 2.2$ & $\pm 2.9$ & $\pm 2.5$ & $\pm 3.5$ & \\ \cline{2-9}
  & $1284.4^{d}$ &  & 0.5 &  &  &  & & \citet{hal03} \\ \cline{2-9}
  & ${1366^{d}}_{-127}^{+199}$ & ${171^{d}}_{-104}^{+564}$ & $0.63_{-0.07}^{+0.09}$ & $35.2_{-3.2}^{+4.8}$ & $175_{-6}^{+5}$ & $63_{-11}^{+12}$ & $61_{-5}^{+5}$ & \citet{gol06,gol07} \\ \cline{2-9}
  & $1278.17^{d}$ & 2013.543 & 0.508 & 101.4 & 356.0 & 246.9 & 58.7 & \citet{sch16} \\
  & $\pm 0.39^{d}$ & $\pm 0.001$ & $\pm 0.001$ & $\pm 0.1$ & $\pm 0.3$ & $\pm 0.1$ & $\pm 0.4$ & \\ \cline{2-9} 
 & $1284.11^{d}$ & 2013.55 & 0.5119 &  &  & 244.75 &  & \citet{kif18} \\
  & $\pm 0.14^{d}$ & $\pm 0.001$ & $\pm 0.0012$ &  &  & $\pm 0.2$ &  & \\ \cline{2-9} 
  & 3.514 & 2013.550 & 0.507 & 102.3 & 177.6 & 244.5 & 58.6 & This work \\
  & $\pm 0.002$ & $\pm 0.003$ & $\pm 0.003$ & $\pm 0.5$ & $\pm 0.4$ & $\pm 0.4$ & $\pm 0.3$ & \\
\hline
73085 & 41.192 & 2014.759 & 0.501 & 205.6 & 22.1 & 142.6 & 127.0 & \citet{cve16} \\  
  & $\pm 0.695$ & $\pm 0.586$ & $\pm 0.038$ & $\pm 9.0$ & $\pm 2.2$ & $\pm 3.8$ & $\pm 1.4$ & \\ \cline{2-9}
  & 43.6 & 2014.66 & 0.475 & 216.4 & 22.6 & 136.9 & 123.9 & This work \\ 
  & $\pm 0.6$ & $\pm 0.06$ & $\pm 0.004$ & $\pm 1.6$ & $\pm 0.5$ & $\pm 1.3$ & $\pm 0.8$ & \\
	\enddata
\end{deluxetable*}

Two independent methods were used to determine the fundamental parameters of the components and systems in this work. The orbital parameters obtained and shown in Table \ref{tab2} and Hipparcos \citep{hippi} and Gaia \citep{gplx} parallaxes were used in the first method. The use of the Hipparcos parallax allows for the comparison of the parameters of objects calculated in this work with published values, and the Gaia parallax - to obtain actual values. The total mass is calculated using 
\begin{equation}
\sum \mathfrak{M}=\frac{(a/\pi)^{3}}{P_{orb}^{2}},
\label{eq2}
\end{equation}
and the uncertainty is given by
\begin{equation}
d\mathfrak{M}=\sqrt{9*\left(\frac{d\pi}{\pi}\right)^2+9*\left(\frac{da}{a}\right)^2+4*\left(\frac{dP_{orb}}{P_{orb}}\right)^2}*\mathfrak{M}.
\label{eq3}
\end{equation}

The apparent magnitudes of the stars from the SIMBAD database, the average magnitude differences of the components in the 550 bandpass from Table \ref{tab1} and Hipparcos and Gaia parallaxes (hereinafter $\pi_{H}$ and $\pi_{G}$) were used in the second method. The calculation of the average magnitude difference in the 550 nm bandpass ($\Delta m_ {550}$) for each object was carried out taking into account all measurements in this filter, including filters with very similar center wavelengths (562 and 545 nm). Since the measurements of the magnitude difference presented in Table \ref{tab1} were obtained by different authors at different telescopes and instruments, the value of the population standard deviation of these measurements was taken as an error. Published and new magnitude differences are significantly differ from each other for HIP~14524 and HIP~73085. To determine the fundamental parameters of these systems, the values obtained at the BTA from 2007 to 2019 were selected, and the population standard deviation described in Section \ref{sec:observations} was used as the measurement error. Only one measurement of $\Delta m_ {550}$ is presented in Table \ref{tab1} for HIP~60444, which was used to determine the parameters (the population standard deviation described in Section \ref{sec:observations} was chosen as an error). This method allows for obtaining of absolute magnitudes of the components, their spectral types and masses, and therefore the estimation of the mass sums. Spectral types and masses were determined based on data from \citet{pec13} for main-sequence stars. Comparison of the results obtained by two methods and using two parallax values allows for estimation of the consistency of a one or another orbital solution.

Table \ref{tab3} lists the fundamental parameters of objects and their components. The columns give the Hipparcos number, the absolute magnitudes of the components ($M_ {A}$ and $M_ {B}$), their spectral types ($Sp_ {A}$ and $Sp_{B}$), the masses of stars ($\mathfrak{M}_{A}$ and $\mathfrak{M}_{B}$), the mass sum of the components defined by orbital parameters by the first method ($\sum \mathfrak{M}$), parallax source, and references. Below, we show the orbits obtained and discuss each system in detail. 

\begin{longrotatetable}
\begin{deluxetable*}{llllllllll}
	\tablenum{3}
	\tablecaption{Fundamental parameters of the objects. \label{tab3}}
	\tablewidth{0pt}
	\tabletypesize{\scriptsize}
	\tablehead{
	\colhead{HIP} & \colhead{$M_{A}$,} & \colhead{$Sp_{A}$} & \colhead{$\mathfrak{M}_{A}$,} & \colhead{$M_{B}$,} & \colhead{$Sp_{B}$} & \colhead{$\mathfrak{M}_{B}$,} & \colhead{$\sum \mathfrak{M}$,} &  \colhead{Parallax}  & \colhead{Reference} \\
	\colhead{ } & \colhead{mag} & \colhead{ } & \colhead{$\mathfrak{M}_{\odot}$} & \colhead{mag} & \colhead{ } & \colhead{$\mathfrak{M}_{\odot}$} & \colhead{$\mathfrak{M}_{\odot}$} & \colhead{source} & \colhead{ } \\
	}
	\decimalcolnumbers
	\startdata
14524 & $7.58 \pm 0.10$ & K5.5V & 0.66 & $8.41 \pm 0.14$ & K8V & 0.59 & $4.96 \pm 3.54$ & Hipparcos & This \\ \cline{2-9}
  & $8.56 \pm 0.10$ & K8V-K9V & 0.56-0.59 & $9.39 \pm 0.10$ & M0.5V & 0.54 & $1.3 \pm 0.2$ & This work & work \\
\hline
16025 & $3.86 \pm 0.16$ & F7 & 1.30 & $5.63 \pm 0.20$ & G8 & 0.97 & $2.12 \pm 0.47$ & Hipparcos & \citet{cve16} \\ \cline{2-10}
  & $4.51 \pm 0.07$ & G0V-G1V & 1.07-1.08 & $6.31 \pm 0.10$ & K2V-K2.5V & 0.76-0.78 & $2.3 \pm 0.5$ &Hipparcos & This \\ \cline{2-9}
  & $3.75 \pm 0.07$ & F6V-F7V & 1.21-1.25 & $5.55 \pm 0.10$ & G9V & 0.9 & $6.5 \pm 1.5$ & Gaia & work \\
\hline
28671 &  &  & 0.9 &  &  & 0.6 &  & \citet{tok16a} & \citet{tok16a} \\ \cline{2-10}
  & $5.61 \pm 0.08$ & G9V & 0.9 & $7.49 \pm 0.12$ & K5V-K5.5V & 0.66-0.68 & $3.8 \pm 1.7$ & Hipparcos & This \\ \cline{2-9}
  & $5.28 \pm 0.08$ & G7V-G8V & 0.94-0.96 & $7.15 \pm 0.12$ & K4.5V-K5V & 0.68-0.71 & $6 \pm 1.5$ & Gaia & work \\
\hline
46199 & $7.11 \pm 0.08$ & K4 & 0.72 & $9.44 \pm 0.36$ & M1 & 0.43 & $0.38 \pm 0.08$ & Hipparcos & \citet{cve16} \\ \cline{2-10}
  & $7.10 \pm 0.07$ & K4.5V & 0.71 & $9.66 \pm 0.10$ & M1V & 0.49 & $0.45 \pm 0.07$ & Hipparcos & This \\ \cline{2-9}
  & $7.38 \pm 0.07$ & K5V & 0.68 & $9.94 \pm 0.10$ & M1.5V-M2V & 0.44-0.47 & $0.30 \pm 0.04$ & Gaia & work \\
\hline
47791 & $1.76 \pm 0.07$ & A4 & 2.10 & $1.76 \pm 0.07$ & A4 & 2.10 & $2.81 \pm 0.29$ & Hipparcos & \citet{cve16} \\ \cline{2-10}
  & $1.87 \pm 0.10$ & A4V-A6V & 1.83-1.9 & $2.29 \pm 0.14$ & A8V-A9V & 1.67 & $3.0 \pm 0.3$ & Hipparcos & This \\ \cline{2-9}
  & $1.70 \pm 0.10$ & A3V-A4V & 1.9-2 & $2.11 \pm 0.14$ & A7V-A8V & 1.67-1.76 & $3.8 \pm 0.1$ & Gaia & work \\
\hline
60444 & $10.85 \pm 0.12$ & M2.5V & 0.4 & $12.39 \pm 0.15$ & M3.5V & 0.26 & $0.64 \pm 0.08$ & Hipparcos & This work \\
\hline
61100 &  &  & $0.67 \pm 0.17$ &  &  & $0.58 \pm 0.1$ &  &  & \citet{are00} \\ \cline{2-10}
  &  &  & 0.75 &  &  & 0.64 &  &  & \citet{hal03} \\ \cline{2-10}
  &  &  & $0.83 \pm 0.02$ &  &  & $0.64 \pm 0.02$ & $1.47 \pm 0.03$ &  & \citet{sch16} \\ \cline{2-10}
  &  &  & $0.834 \pm 0.017$ &  &  & $0.640 \pm 0.011$ &  &  & \citet{kif18} \\ \cline{2-10}
  & $6.24 \pm 0.08$ & K2V-K2.5V & 0.76-0.78 & $8.20 \pm 0.12$ & K6.5V-K7.5V & 0.61-0.64 & $1.41 \pm 0.03$ & Gaia & This work \\
\hline
73085 & $7.35 \pm 0.63$ & K5 & 0.65 & $7.74 \pm 0.66$ & K6 & 0.60 & $1.91 \pm 1.67$ & Hipparcos & \citet{cve16} \\ \cline{2-10}
  & $7.10 \pm 0.11$ & K4V-K4.5V & 0.71-0.72 & $8.71 \pm 0.15$ & K9V & 0.56 & $2.0 \pm 1.7$ & Hipparcos & This \\ \cline{2-9}
  & $7.45 \pm 0.11$ & K5V-K5.5V & 0.66-0.68 & $9.06 \pm 0.15$ & M0.5V-M0V & 0.54-0.55 & $1.23 \pm 0.08$ & Gaia & work \\
	\enddata
\end{deluxetable*}
\end{longrotatetable}

\begin{figure*}[ht!]
	\plottwo{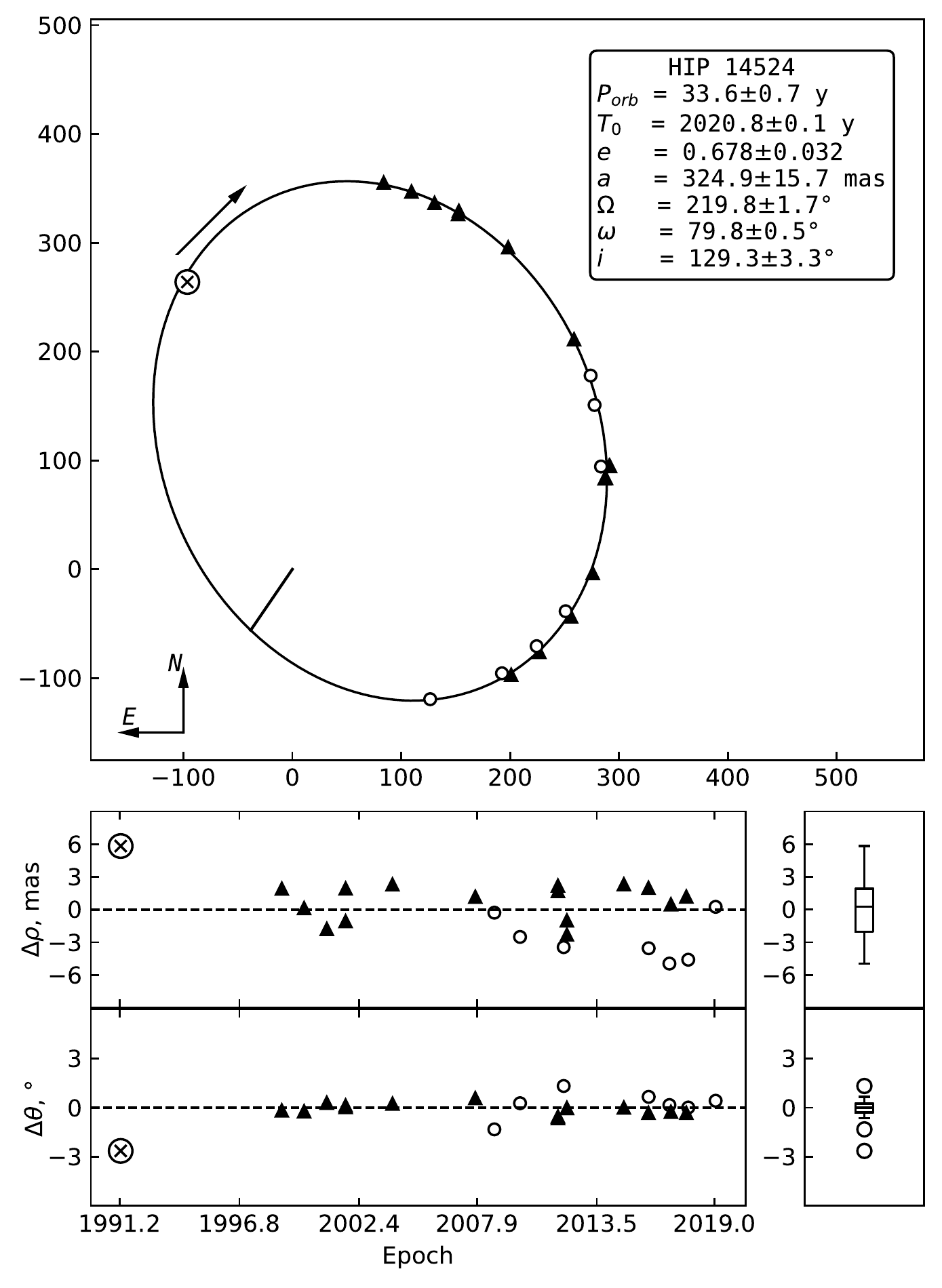}{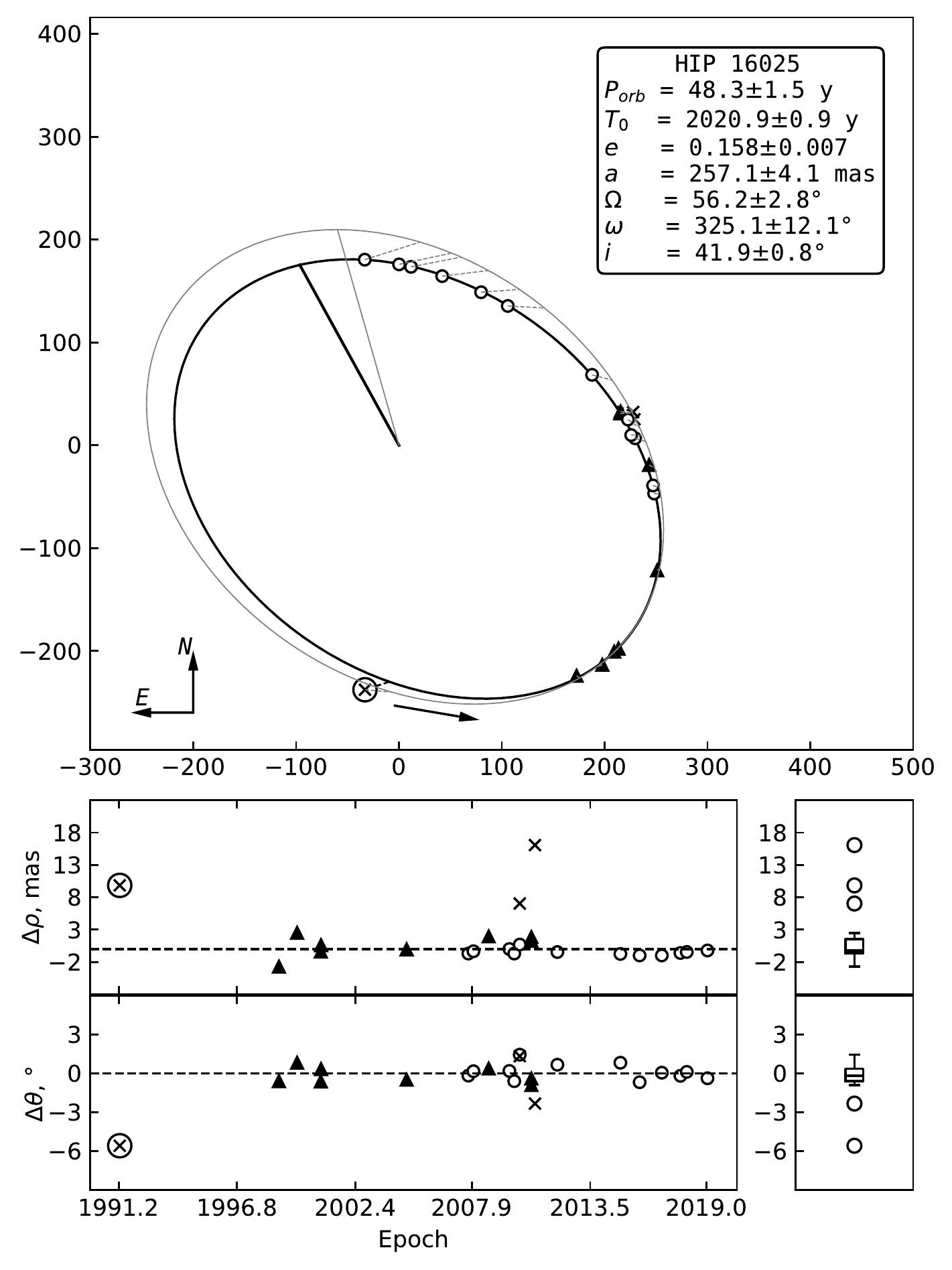}
	\caption{Orbital solutions for HIP~14524 and HIP~16025. The orbit by \citet{cve16} is marked with gray and the orbit constructed in this work is black. Triangles correspond to the published data; open circles - data obtained in this study; crosses - data with large residuals; a point placed in the large circle is the first measurement for system. The arrow shows the direction of motion of the secondary. $\Delta \rho$ and $\Delta \theta$ are residuals of angular separation and position angle showing difference between the observed and model value. The dashed line on the residuals plots indicates the orbital solution. Box plots display the distribution of data based on a five-number summary (“minimum”, first quartile (Q1), median, third quartile (Q3), and “maximum”) and outliers.  \label{fig3}}
\end{figure*}

{\bf HIP~14524} ($03^{h} 07^{m} 33\fs78 -03\degr 58\arcmin 17\farcs14$; MCC~413) is a binary consisting of components of the spectral types K6 and K7 with absolute magnitudes of $M_{A} = 7.6$ mag and $M_{B} = 8.0$ mag \citep{bali02}. The orbital solution for this system was obtained for the first time and is shown in Figure \ref{fig3}. The calculation includes 16 previously published measures and the 7 new measurements that appear in Table \ref{tab1}. Their residuals regarding the orbital solution were $\Delta \rho = 2.6$ mas and $ \Delta \theta = 0\fdg8$. However, the measurement of the Hipparcos mission, which has the biggest discrepancies in both $\rho$ and $\theta$, contributes to these values (and is marked with a cross in Figure \ref{fig3}). With its exclusion, the residuals are $\Delta \rho = 2.3$ mas and $\Delta \theta = 0\fdg5$, and when using only new measurements, they are $\Delta \rho = 0.9$ mas and $\Delta \theta = 0\fdg4$. Only the Hipparcos parallax $\pi_{H} = 18.29 \pm 4.26$ mas is presented in the SIMBAD database for HIP~14524, and the value of the magnitude difference (as stated earlier) is $0.82 \pm 0.10$ mag. A comparison of the mass sums (Table \ref{tab3}) calculated by the two methods shows that the value obtained by the orbital solution is obviously overestimated and has a large error of about 71\% due to the low accuracy of the Hipparcos parallax (the error of the value is 23\%). However, in Figure \ref{fig3} it can be seen that the orbit fits the observational data well. Using the spectral type of the system from the SIMBAD database (K7V), the corresponding absolute magnitude $M_ {V} = 8.15$ mag \citep{pec13} and the magnitude difference, an implied parallax of $\pi = 28.7 \pm 0.5$ mas is calculated. The fundamental parameters calculated using this parallax are in good agreement with each other. If this analysis is correct, then the stars are of later spectral types than \citet{bali02} assumed.

{\bf HIP~16025} ($03^{h} 26^{m} 21\fs02 +35\degr 20\arcmin 26\farcs18$; HD~21183) is a binary with components of the F8 ($M_{A} = 4.1$ mag) and K0 ($M_{B} = 5.8$ mag) spectral types \citep{bali02}, effective temperature $T_{eff}=5703$ K, luminosity $L=1.72~L_{\odot}$ and infrared excess $E_{IR}=1.007$ \citep{mcd12}. The infrared excess indicates the presence of a secondary. Using the Hipparcos parallax and $\Delta m = 1.77 \pm 0.12$ mag \citet{cve16} determined spectral types of components, their masses and the total mass of the system, which are given in Table \ref{tab3} for subsequent comparison, and the dynamical parallax $\pi_ {dyn} = 15.6 \pm 0.44$ mas.

Analysis of new speckle interferometric data made it possible to increase the number of measurements for constructing the orbit by 13 and to improve the already known orbital solution by \citet{cve16} (Figure \ref{fig3}, right panel). The residuals of measurements were $1\fdg3$ for $\theta$ and 4 mas for $\rho$. However, they are overestimated due to the contribution of measurements, which are marked with crosses in Figure \ref{fig3}. After their exclusion, the residuals are $\Delta \rho = 1.3$ mas and $\Delta \theta = 0\fdg6$, and when using only new measurements, they are $\Delta \rho = 0.7$ mas and $\Delta \theta = 0\fdg5$. The mass sums obtained from the parallaxes $\pi_{H} = 14.76 \pm 1.06$ mas and $\pi_{G} = 10.3951 \pm 0.7421$ mas are very different from each other. Such a big difference occurs due to an error in the parallax of the Gaia mission, since at scales less than $0\farcs5$ this data release rarely resolves objects as double ones \citep{gplx}, while in Equation \ref{eq1} this value is raised to the third power. To calculate the parameters of the components by the second method, the value of the average magnitude difference $\Delta m = 1.80 \pm 0.07$ mag was used (where 13 measurements are from Table \ref{tab1}). The masses of stars calculated by the second method are in good agreement with the mass sum obtained from the orbital parameters with the Hipparcos parallax. The orbit constructed in this study (Figure \ref{fig3}, right panel) fits most of the measurements, and the discrepancy in the mass sums is probably due to problems with the currently known parallaxes of the Gaia mission.

{\bf HIP~28671} ($06^{h} 03^{m} 14\fs86 +19\degr 21\arcmin 38\farcs7$, J2000; HD~250792) is a triple system of spectral type G, consisting of a binary ($BD~+19\degr1185A$) and a physically bound third component ($BD~+19\degr1185B$) at a distance of $\rho \approx 6\farcs9$ \citep{rast07, hart12}. The orbital period of the system was initially estimated as $\sim 10$ years \citep{hart12}, the distance to the object was $d = 54.3 \pm 20.9$ pc, and the trigonometric parallax $\pi = 31.4 \pm 7.1$ mas \citep{khov13}. This distance value was obtained taking into account the Lutz-Kelker correction \citep{lut73}; therefore, it differs from the preliminary estimates calculated as $d=1/\pi$. \citet{khov13} also note that mutual orbital motions of the components of double systems might affect the results of parallax determinations. Based on the data from the MILES, CFLIB, and ELODIE, \citet{pru11} determined the atmospheric parameters of HIP~28671 as $T_{eff} = 5554 \pm 42$ K, $log~g = 4.33 \pm 0.07$, and $[Fe/H] = -1.01 \pm 0.05$ dex, which is consistent with the results by \citet{schu12} ($T_{eff} = 5489 \pm 148$ K, $log~g = 4.47 \pm 0.05$, and $[Fe/H] = -1.01 \pm 0.02$ dex). \citet{tok16a} constructed a preliminary orbit of the object and iteratively determined the masses of the components and dynamical parallax $\pi_ {dyn} = 18.0$ mas. The values obtained by \citet{tok16a} analysis are shown in Tables \ref{tab2} and \ref{tab3}.

\begin{figure*}[ht!]
	\plottwo{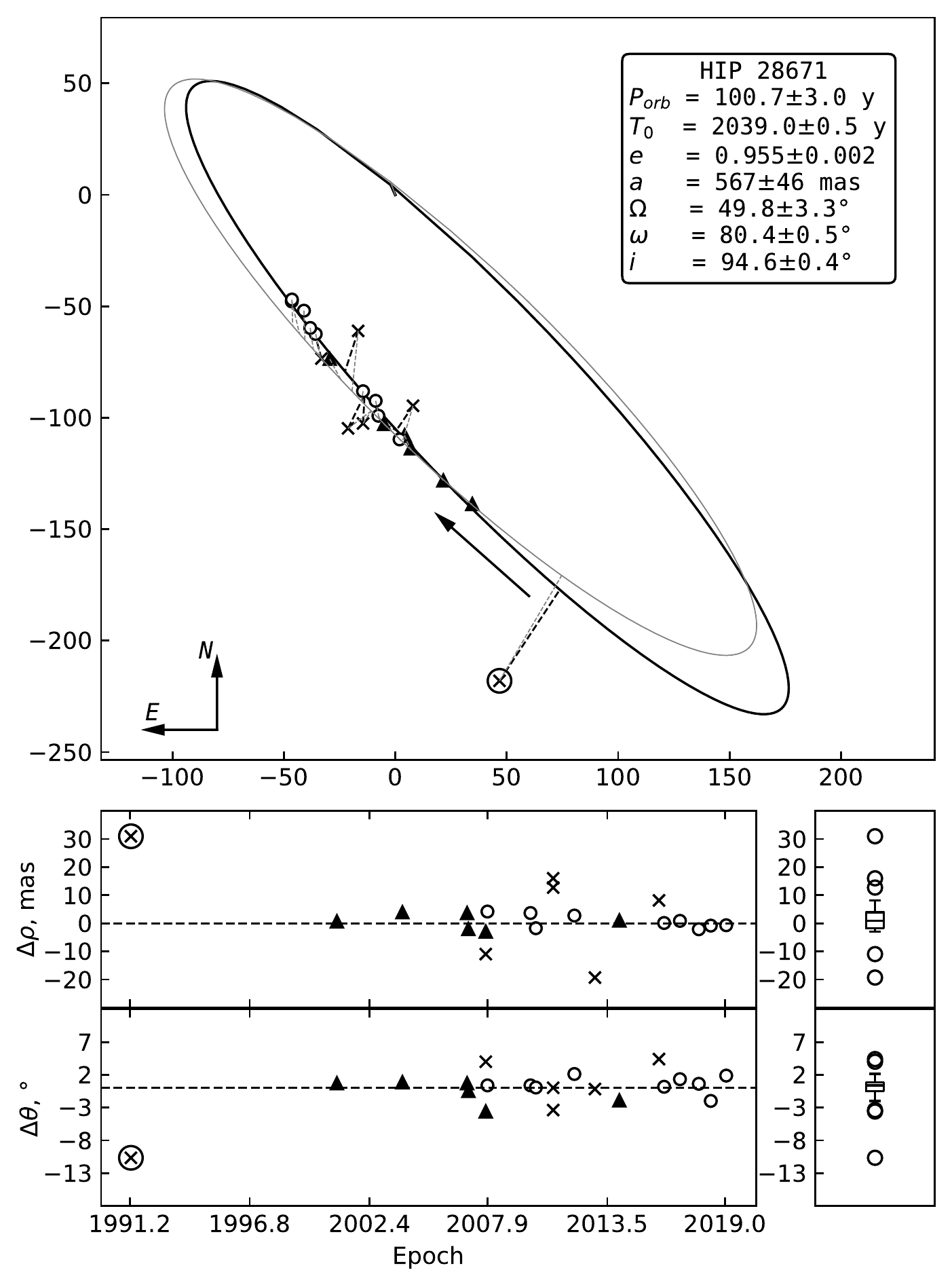}{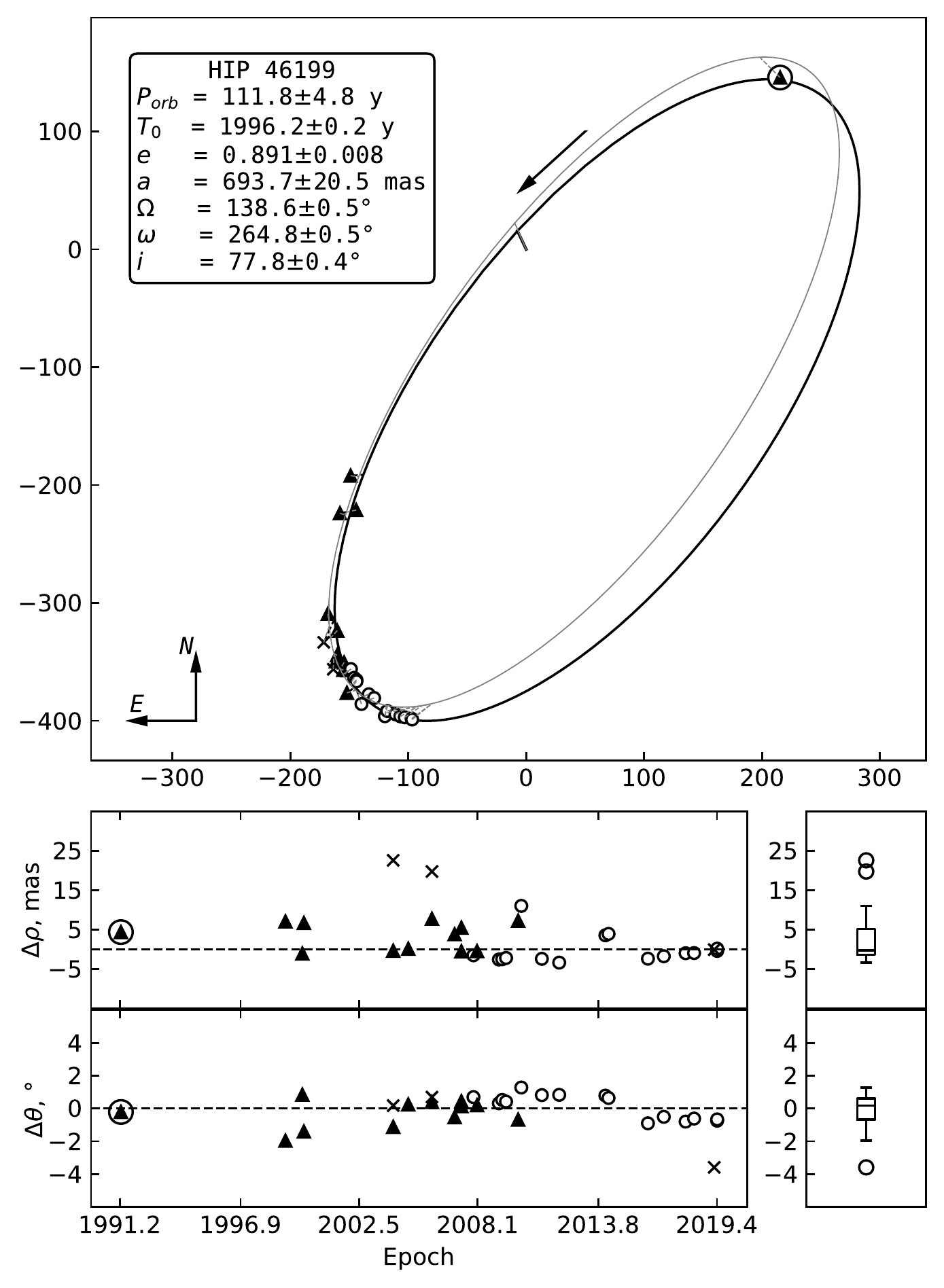}
	\caption{Orbital solutions for HIP~28671 and HIP~46199. The orbit by \citet{tok16a} is shown in gray on the right panel and by \citet{cve16} on the left. All other designations are described in the caption of Figure \ref{fig3}.  \label{fig4}}
\end{figure*}

The orbital parameters of HIP~28671 can only be refined due to the expected long orbital period of the object and the small number of measurements (12 published and 9 new) covering only a fifth of the preliminary orbit (Figure \ref{fig4}, left panel). If all measurements are used to construct the orbit, then the residuals are $\Delta \rho = 9.8$ mas and $\Delta \theta = 3\fdg2$. With the exclusion of data that are not in good agreement with the orbital solution (marked with crosses in Figure \ref{fig3}), the residuals are $\Delta \theta = 1\fdg4$ and $\Delta \rho = 2.5$ mas. If only new measurements are used, then the residuals are $\Delta \theta = 1\degr$ and $\Delta \rho = 1.7$ mas. The study made it possible to refine the orbital solution constructed by \citet{tok16a}, and to compare the mass sums obtained for both solutions. Based on the parallaxes of $\pi_{H} = 16.81 \pm 2.04$ mas and $\pi_{G} = 14.4036 \pm 0.1174$ mas, the mass sums of the components were obtained for the orbital solution by \citet{tok16a} ($\sum \mathfrak{M}_{H} = 1.9 \mathfrak{M}_{\odot}$ and $\sum \mathfrak{M}_{G} = 3.03 \mathfrak{M}_{\odot}$) and for the orbit presented here in Table \ref{tab3}. Using the parallax in \citet{khov13}, the mass sums are $\sum \mathfrak{M}_{khov} = 0.3 \mathfrak{M}_{\odot}$ \citep{tok16a} and $\sum \mathfrak{M}_{khov} = 0.58 \pm 0.42 \mathfrak{M}_{\odot}$ (this work). The mass sums using this parallax for both orbital solutions are implausibly low, therefore, this parallax was not used in a further study. It should be noted that the low accuracy of the values obtained from the orbital parameters is due to large inaccuracies in the determination of the semimajor axis and the orbital period. The high proper motion of the object, the small number of measurements, and differences in the parallaxes for HIP~28671 currently do not allow for reliable determination of the mass sum of the components. Using the average magnitude difference of the two stars ($\Delta m_{550} = 1.87 \pm 0.08$ mag, derived from 5 measurements in Table \ref{tab1}), the characteristics of the components presented in Table \ref{tab3} were calculated. Despite the best description of the observational data by the new orbital solution, the mass sum, determined according to the data by \citet{tok16a} and the Hipparcos parallax, is closest to the estimated mass sum obtained using the magnitude differences. Obviously, further monitoring of HIP~28671 is required in order to cover more of the orbital period.

{\bf HIP~46199} ($09^{h} 25^{m} 10\fs78 +46\degr 05\arcmin 53\farcs65$; HD~81105) consists of stars of spectral types K4 and M1 \citep{cve16}. Based on the published positional parameters, \citet{cve16} found an orbital solution for this system and determined parameters of the components (using  $\Delta m = 2.33 \pm 0.35$ mag) and $\pi_{dyn} = 25.49 \pm 2.78$ mas. The analysis of new observational data for HIP~46199 obtained at the 6 m BTA of the SAO RAS from 2007 to 2019 increased the number of measurements of positional parameters from 15 to 31 (where the new orbit is shown in the right panel of Figure \ref{fig4}). The residuals of measurements are $1\fdg0$ for $\theta$ and 6.7 mas for $\rho$. After excluding the contribution of measurements with large residuals (in Figure \ref{fig4}, these are marked with crosses), these values are $1\degr$ and 5.6 mas, and when using only new measurements, they are $\Delta \rho = 2.6$ mas and $\Delta \theta = 0\fdg8$. The mass sums obtained using both parallaxes, $\pi_{H} = 38.9 \pm 1.21$ mas and $\pi_{G} = 44.3975 \pm 0.8210$ mas, appearing in Table \ref{tab3} are obviously too low. Most likely this is due to the low accuracy of the positional parameters of the Hipparcos mission (for some systems, these measurements have the largest residuals). If this measurement is excluded from the study, only a small part of the arc remains for fitting, which leads to many orbital solutions with similar residuals. The above indicates the need for further monitoring of HIP~46199. There is no information in the SIMBAD database about the apparent magnitude of the system in the V band, therefore, to obtain this value, we used photometric ratios and conversion rate from \citet{carr18}. As a result, using the values of $m_{V} = 9.05 \pm 0.04$ mag and $\Delta m_{550} = 2.56 \pm 0.06$ mag obtained from the 6 measurements in Table \ref{tab1}, the masses of the components were obtained.

{\bf HIP~47791} ($09^{h} 44^{m} 36\fs55 +64\degr 59\arcmin 02\farcs79$; HD~83962) is a binary of the spectral type F3Vn \citep{cowl76, mcal93} with $v \cdot \sin i = 145$ km $s^{-1}$  \citep{dan72}. \citet{mcd12} gives the effective temperature of an object $T_{eff}=6583$ K, luminosity $L=24.42~L_{\odot}$ and infrared excess $E_{IR}=1.429$. Based on the published data, \citet{cve16} calculated the orbital parameters of HIP~47791 shown in Table \ref{tab2} and a number of characteristics of the components (in Table \ref{tab3}), as well as $\pi_{dyn} = 9.61 \pm 0.08$ mas. 

\begin{figure*}[ht!]
	\plottwo{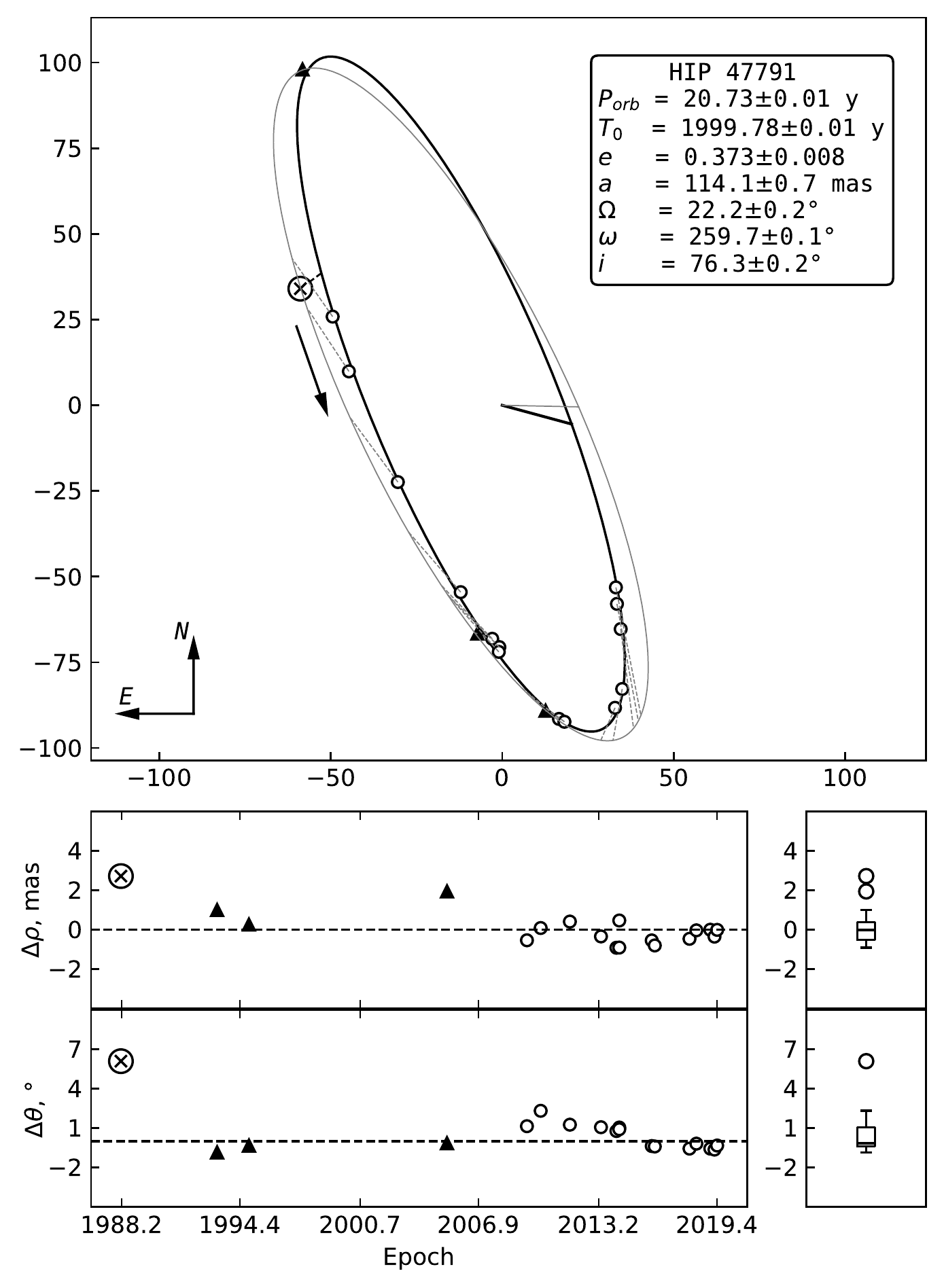}{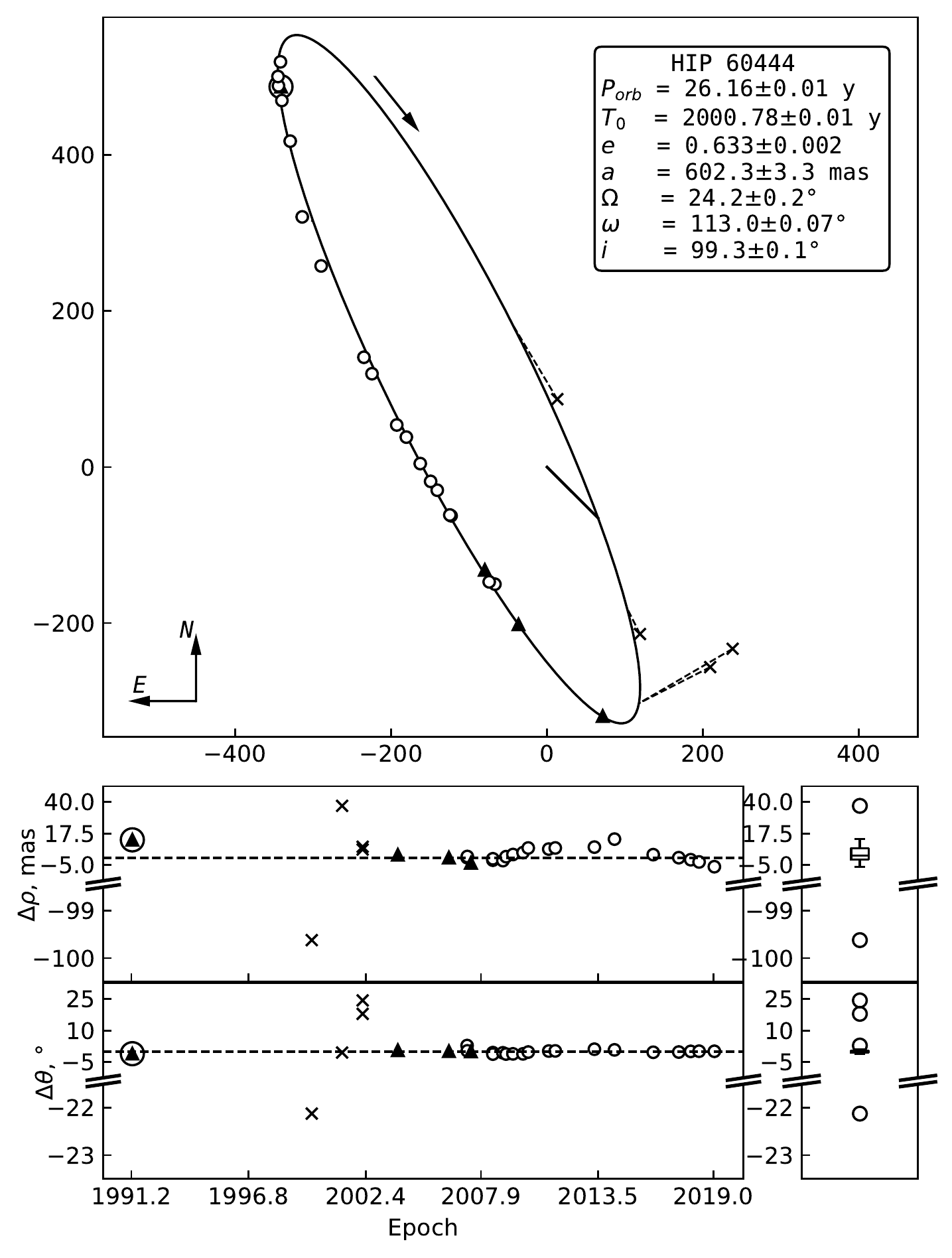}
	\caption{Orbital solutions for HIP~47791 and HIP~60444. The orbit by \citet{cve16} is shown in gray on the left panel. All other designations are described in the caption of Figure \ref{fig3}. \label{fig5}}
\end{figure*}

The analysis of 4 published and 14 new speckle interferometric measurements made it possible to improve the orbit of HIP~47791. The result is shown in the left panel of Figure \ref{fig5}. The residuals of measurements were $1\fdg7$ for $\theta$ and 1 mas for $\rho$. The measurements made by \citet{mcal93} give the largest residuals; without these, the residual in $\theta$ is reduced to 0.9 degrees and in $\rho$, the value is reduced to 0.7 mas. Residuals of new measurements are $\Delta \rho = 0.4$ mas and $\Delta \theta = 0\fdg9$. When constructing the orbit, \citet{cve16} suggested that the position angle of the measurement by \citet{bali13} should be modified by $180\degr$, but in our analysis that appears not to be needed. The parallaxes of Hipparcos and Gaia for HIP~47791 are equal to $\pi_{H} = 10.48 \pm 0.36$ mas and $\pi_{G} = 9.6586 \pm 0.1059$ mas, respectively. The parameters obtained using the orbit calculated here and the magnitude difference between the components $\Delta m_{550} = 0.42 \pm 0.01$ mag (including 12 measurements from Table \ref{tab1}) are in good agreement with each other, as can be seen in Table \ref{tab3}. In the work by \citet{cve16}, the magnitude difference of the components was taken equal to zero, since there were no estimates of this value in the literature. Using the magnitude difference in this study indicates that the secondary has a later spectral type.

{\bf HIP~60444} ($12^{h} 23^{m} 33\fs18 +67\degr 11\arcmin 17\farcs91$, J2000; G~237-64) consists of two components of M4 spectral types with absolute magnitudes $M_{V}^{A} = 11.4$ mag and $M_{V}^{B} = 11.6$ mag \citep{bali04}. The orbital solution for this system was obtained for the first time and is shown in Figure \ref{fig5}. It is based on 26 measurements (8 published and 18 new), and the residuals of measurements were $\Delta\rho = 21.6$ mas and $\Delta\theta = 7\fdg4$. With the exclusion of measurements marked with a crosses in Figure \ref{fig5}, the residuals are $\Delta\rho = 5.1$ mas and $\Delta\theta = 0\fdg8$. Residuals of new measurements are $\Delta\rho = 4.3$ mas and $\Delta\theta = 0\fdg8$. Only the Hipparcos parallax $\pi_{H} = 77.54 \pm 3.23$ mas is presented in the SIMBAD database for HIP~60444, and the value of the magnitude difference used in our analysis is $\Delta m_{550} = 1.54 \pm 0.10$ mag as stated earlier. The mass sums calculated by the two methods and appearing in Table \ref{tab3} are in good agreement with each other, which indicates the high precision of the orbital solution.

{\bf HIP~61100} ($12^{h} 31^{m} 18\fs91 +55\degr 07\arcmin 08\farcs28$, J2000; NO~UMa) belongs to the Ursa Major moving group \citep{kin03, amm09}. The components have spectral types $K2.0V \pm 0.5$ (NO~UMa~A) and $K6.5V \pm 0.5$ (NO~UMa~B) \citep{sch16}. In addition to estimates of the positional and orbital parameters of the system appearing in Tables \ref{tab1} and \ref{tab2}, respectively, the following parameters are available in the literature (and some appear in Table \ref{tab3}): $q = 0.86$ \citep{hal03}; $T_{eff} = 5000 \pm 100$ K, $log~g = 4.6 \pm 0.2$, $[Fe/H] = -0.13 \pm 0.07$ dex, $\mathfrak{M}=0.83 \pm 0.05 \mathfrak{M}_{\odot}$, $R = 0.81 \pm 0.07 R_{\odot}$ and $d = 22.3 \pm 5.3$ pc \citep{amm09}; $T_{eff,A} = 5010 \pm 50$ K, $T_{eff,B} = 4140 \pm 30$ K and $d = 25.87 \pm 0.02$ pc \citep{sch16}; and an excess of helium was also detected \citep{kif18}. In paper by \citet{kif18}, parameters were determined with high accuracy based on spectroscopic and speckle interferometric data. As a result, 5 sets of orbital parameters are presented in the literature for this object in Table \ref{tab2}, but it should be noted that some of the values are not consistent with each other.

\begin{figure*}[ht!]
	\plottwo{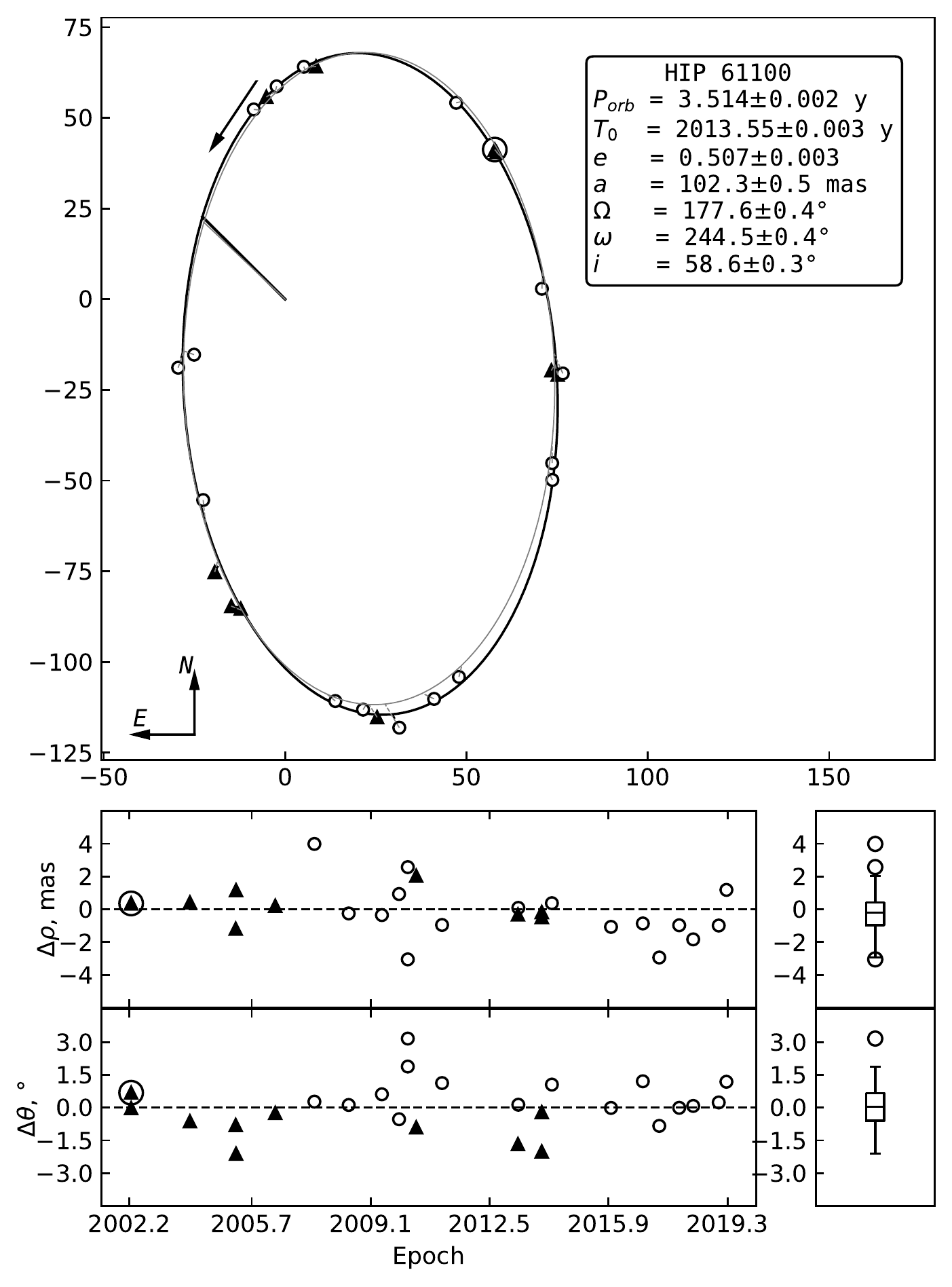}{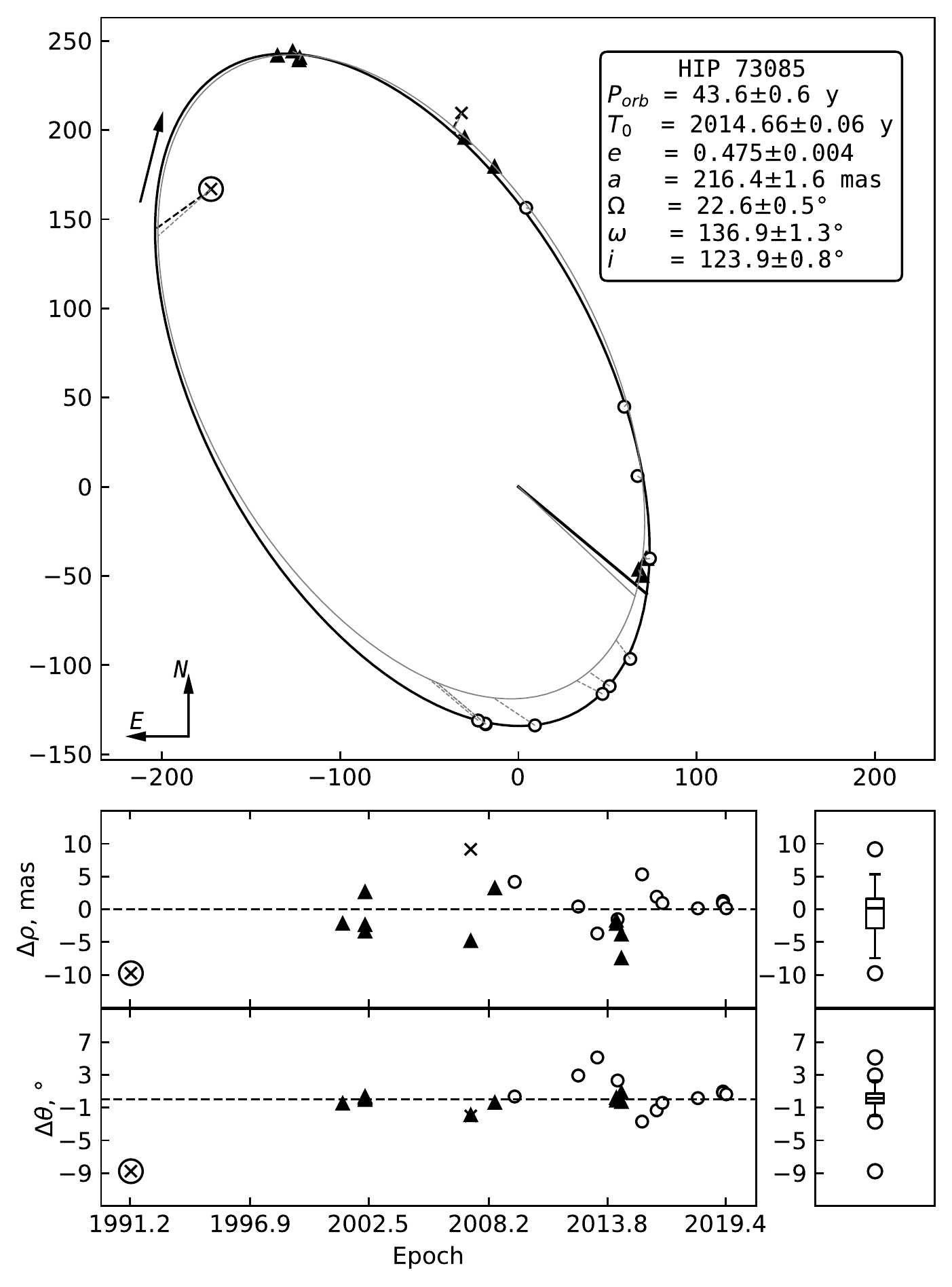}
	\caption{Orbital solutions for HIP~61100 and HIP~73085. The orbit by \citet{sch16} is shown in gray on the left panel and by \citet{cve16} on the right panel. All other designations are described in the caption of Figure \ref{fig3}. \label{fig6}}
\end{figure*}

For this system, 16 new measurements appear in Table \ref{tab1} in addition to the 10 already available in the literature. The orbital solution found for HIP~61100 in this work made it possible to refine the already known orbital parameters that we present in Table \ref{tab2} and fit the observational data. Improved orbit and the orbit constructed according to the parameters by \citet{sch16} are shown in Figure \ref{fig6}. It should be noted that the longitude of the ascending node $\Omega$ from the paper by \citet{sch16} has been changed by $180\degr$. The residuals of the measurements for $\rho$ and $\theta$ are 1.5 mas and $1\fdg4$, respectively. When using only new measurements, they are $\Delta \rho = 1.4$ mas and $\Delta \theta = 1\fdg3$. The Gaia and Hipparcos parallaxes for HIP~61100 have very close values of $\pi_{H} = 39.84 \pm 1.07$ mas \citep{hippi} and $\pi_{G} = 39.4502 \pm 0.2049$ mas \citep{gplx}, therefore, it makes sense to provide calculations based on Gaia parallax. To calculate the absolute magnitudes of the stars, we used the average magnitude difference $\Delta m_{550} = 1.96 \pm 0.02$ mag (calculated from 6 measurements from Table \ref{tab1}). The mass sums calculated by the two methods in this study are in excellent agreement with each other.

{\bf HIP~73085} ($14^{h} 56^{m} 12\fs87 +17\degr 44\arcmin 53\farcs48$; MCC~727) is a binary consisting of main-sequence stars. For this object \citet{cve16} obtained orbital parameters and characteristics of the system that are shown in Tables \ref{tab2} and \ref{tab3}, and a dynamical parallax of $\pi_{dyn} = 16.63 \pm 1.12$ mas. The orbit was improved using 12 published and 11 new measurements, which made it possible to fit the observational data better, shown in the right panel of Figure \ref{fig6}. The measurement residuals were $2\fdg4$ for $\theta$ and 4 mas for $\rho$. Measurements that are very different from the model solution contribute to these values (in Figure \ref{fig6}, marked with crosses), so the final estimates of the residuals for $\rho$ and $\theta$ are 3 mas and $1\fdg5$, respectively. Residuals for new measurements are $\Delta \rho = 1.7$ mas and $\Delta \theta = 1\fdg5$. As a result, the orbit obtained here fits the observational data better. The parallaxes for HIP~73085 are $\pi_{H} = 13.89 \pm 4.00$ mas and $\pi_{G} = 16.3101 \pm 0.2908$ mas, and the value of the magnitude difference of stars is $\Delta m_{550} = 1.61 \pm 0.10$ mag (as noted earlier). The high value of the error in determining of the mass sum using the Hipparcos parallax is due to its large uncertainty. The characteristics of the components and the system, calculated by two methods and using different parallaxes, are in good agreement with each other (Table \ref{tab3}), indicating the high precision of the orbital solution obtained.

\subsection{Grading the Orbits}

Since the measurements cover less than half of the orbital period and further monitoring of HIP~28671 is required, the orbital solution for this system is "preliminary" (Grade 4). "Reliable" (Grade 3) orbits of HIP~14524, HIP~16025, and HIP~46199 are obtained in the cases where at least half of the orbit is defined, but observational data (for example, Hipparcos mission measurements) leave the possibility of inaccuracies in the orbital parameters. The orbits of HIP~60444 and HIP~73085 are "good" (Grade 2) - the observational data correspond to different phases and cover more than half of the orbital period, which allows for an accurate fitting of the orbit; probably, taking into account the positional parameters of systems in the missing phases, the orbital solution will be improved. It should be noted that more than one orbital period has passed since the beginning of the observations of the HIP~47791 and HIP~61100 and the observations are perfectly fitted by the obtained orbital solutions, no significant improvements are expected in the orbital parameters obtained in this work, therefore the orbits of these objects are classified as "accurate" (Grade 1).

We would like to draw attention to the low residuals of positional parameters (an average of $1\degr$ on $\theta$ and 2.7 mas on $\rho$), which are indicators of the high precision of the construction of the orbits. This fact indicates the high accuracy of determining the orbital parameters of the objects under study (Table \ref{tab2}) and the validity of the long-term monitoring of binary and multiple systems performed in the group of high-resolution methods in astronomy of the SAO RAS.

\section{Conclusions} \label{subsec:conclusions}

This study has used speckle interferometric observations obtained at the BTA of the SAO RAS over the past 12 years, in order to significantly increase the number of measurements of positional parameters to construct orbits. The results of the analysis of new speckle interferometric data obtained from 2007 to 2019 at 6 m telescope made it possible to obtain orbital solutions for the HIP~14524 and HIP~60444 for the first time and improve the orbits of HIP~16025, HIP~28671, HIP~46199, HIP~47791, HIP~61100 and HIP~73085. Using the criteria by \citet{wor83}, a qualitative classification of the obtained orbital solutions was carried out.

Based on the orbital elements of the objects under study and estimates of the magnitude differences $\Delta m$, the total masses, absolute magnitudes, and spectral types of components were calculated. The mass sums were determined with error of 15\% using a parallax from this work for HIP~14524 (a high error in determining the semi-major axis of this system should be noted), 15\% and 13\% for HIP~46199 using the Hipparcos and Gaia parallaxes, respectively, 10\% (Hipparcos parallax) and 3\% (Gaia parallax) for HIP~47791, 12\% (Hipparcos parallax) for HIP~60444, 2\% (Gaia parallax) for HIP~61100 and 85\% (Hipparcos parallax) and 6\% (Gaia parallax) for HIP~73085. It should be noted that the large uncertainties of mass sums of some systems are due to the low accuracy of the parallaxes used and the semi-major axes, in some cases. Further monitoring of binaries (in particular HIP~16025 and HIP~28671) and future data releases of Gaia will make it possible to construct orbits better, improve known orbital solutions, and also clarify the problem of the mismatch of the mass sums of components determined using the Hipparcos and Gaia parallaxes.

\acknowledgments

The reported study was funded by RFBR, project number 20-32-70120. The work was performed as part of the government contract of the SAO RAS approved by the Ministry of Science and Higher Education of the Russian Federation. The authors would like to thank the anonymous referee for constructive comments that helped to improve the content and clarity of this paper. This research has made use of the SIMBAD database, operated at CDS, Strasbourg, France. This work has made use of data from the European Space Agency (ESA) mission Gaia (\url{https://www.cosmos.esa.int/gaia}), processed by the Gaia Data Processing and Analysis Consortium (DPAC, \url{https://www.cosmos.esa.int/web/gaia/dpac/consortium}). Funding for the DPAC has been provided by national institutions, in particular the institutions participating in the Gaia Multilateral Agreement.

\bibliography{8orbits}{}
\bibliographystyle{aasjournal}



\end{document}